\documentclass[final,3p,11pt]{elsarticle}
\usepackage{geometry}
\usepackage{graphicx}
\usepackage{amsmath}
\usepackage{xcolor}
\usepackage{amsthm}
\usepackage{amssymb}
\usepackage{lineno}
\usepackage{bm}
\usepackage{multirow}
\usepackage{subcaption}
\usepackage{threeparttable}
\usepackage{todonotes}
\usepackage[linesnumbered,ruled,vlined,longend]{algorithm2e}
\makeatletter
\def\ps@pprintTitle{%
  \let\@oddhead\@empty
  \let\@evenhead\@empty
  \let\@oddfoot\@empty
  \let\@evenfoot\@oddfoot
}
\makeatother
\begin{document}

\begin{frontmatter}

\title{\textbf{Bayesian Methods in Automated Vehicle's Car-following Uncertainties: Enabling Strategic Decision Making}}
\author[add1]{Wissam Kontar}
\author[add1]{Soyoung Ahn\corref{cor1}}
\address[add1]{Civil and Environmental Engineering, University of Wisconsin-Madison, USA}
\cortext[cor1]{Corresponding author: sue.ahn@wisc.edu}

\begin{abstract}
This paper proposes a methodology to estimate uncertainty in automated vehicle (AV) dynamics in real time via Bayesian inference. Based on the estimated uncertainty, the method aims to continuously monitor the car-following (CF) performance of the AV to support strategic actions to maintain a desired performance. Our methodology consists of three sequential components: (i) the Stochastic Gradient Langevin Dynamics (SGLD) is adopted to estimate parameter uncertainty relative to vehicular dynamics in real time, (ii) dynamic monitoring of car-following stability (local and string-wise), and (iii) strategic actions for control adjustment if anomaly is detected. The proposed methodology provides means to gauge AV car-following performance in real time and preserve desired performance against real time uncertainty that are unaccounted for in the vehicle control algorithm. 
\end{abstract}

\end{frontmatter}

%%%%%%%%%%%%%%%%%%%%%%%%%%%%%%%%%%%%%%%%%%%%%%%%%%%%%%%%%%%%%%%%%%%%%%%%%%%%%%%%%%%%%%

\section{Introduction}\label{S:1}

Skepticism toward Autonomous Vehicle's (AVs') ability to coexist in our transportation system in a safe, effective, and desirable manner has been a momentous barrier to the market deployment of AVs. A critical element in the development and deployment of AVs is the design of car-following (CF) controllers capable of producing desirable performance in real-world settings. Ideally, a CF control system would effectively and safely handle the longitudinal maneuvers of the vehicle at every encounter it faces. However, designing and training such a controller requires enormous data, testing, and experimentation that covers all possible driving scenarios/encounters. In other words, it requires us to have a perfect understanding of the environment these AVs would be operating under. Clearly, this is very challenging and, possibly, unattainable. AVs are likely to encounter unseen scenarios and be exposed to exogenous and endogenous uncertainties in the physical world. The sources of exogenous and endogenous uncertainties are vast and roughly classified into \citep{macfarlane2016addressing,yao2020control,katrakazas2015real}: (i) vehicular and system dynamics (e.g., vehicle condition, road gradient, aerodynamic drag force, external loads, transmission, brake, the performance of the engine, etc.), (ii) environmental conditions (snow, dust, wind, wet conditions, etc.), and (iii) situational detection (e.g., sensor/measurement errors, radar errors, vehicle speed fluctuations, vehicle localization, communication latency, etc.). All these types of uncertainties can hinder desirable performance (e.g., stability). Yet, a major challenge lies in the complexity of integrating these uncertainties into the control system and the design of the AV. For instance, it is often hard to formulate an analytical representation of these uncertainties and quantify their impact on performance.
Additionally, even when these uncertainties can be modeled through parameterization, the complexity of formulation increases significantly. The caveat, however, is that such uncertainties are most influential on the CF control as they can alter the desired performance, resulting in actuation lag and a mismatch between demanded acceleration and realized acceleration of the AV. Such behavior has been shown to impact the local/string stability of the vehicle and the traffic system as a whole \citep{yao2020control,kontar2021multi,zhou2020stabilizing,zhou2019robust}.

Guided by empirical experimentation, analytical analysis, and commercial product investigation (e.g., factory ACC vehicles and current self-driving technologies), the literature has recently given specific interest to how vehicular dynamics impact the performance of the vehicle. A typical CF controller consists of an upper-level and lower-level control. The upper-level functions as a planner that receives sensor data (on distance, velocity, acceleration) and sends commands to the lower-level to execute (i.e., braking, accelerating, etc.). Notably, there could be a discrepancy between the commanded action (upper-level) and the executed action (lower-level). This has been shown to be the reality in real-life driving conditions. The assumption of perfect execution of commands by low-level controllers has been shown to be unrealistic, with significant implications on local/string stability and overall performance \citep{zhou2019robust,gunter2019model,li2020trade,zhou2019distributed,wang2018delay,zhou2017parsimonious,yi2001vehicle,zhou2019robust,shi2021empirical,yi2001vehicle,zhou2022significance}. A recent paper by \citep{zhou2022significance} investigates the significance of low-level control for ACC vehicles on string stability. Their theoretical and empirical investigations connect disturbances' frequency and amplifications to low-level control functions.

The exact factors that cause a discrepancy between demanded  (by upper-level) and executed actions (by lower-level) are hard to pinpoint due to the complexity of behavior, non-linearity, and high dimensionality of such factors (multiple exogenous and endogenous factors play a role here). However, significant attention has been given to uncertainties impacting vehicular dynamics. Some limited efforts to deal with such uncertainties have led to the development of robust control methods that adjust system states for uncertain vehicular dynamics. Most notably, the General Longitudinal Vehicle Dynamics (GLVD) model is considered in robust control frameworks to formalize such uncertainties. The GLVD model parameterizes two key uncertainties in vehicular dynamics: actuation lag and the ratio of demanded acceleration that can be realized. The basic idea is to acknowledge that lower-level controllers are not perfectly able to execute the demanded acceleration. Thus the controller adjusts its acceleration according to the GLVD equation. The GLVD parameters are shown to greatly influence CF performance and local/string stability \citep{li2018robust,yi2001vehicle,wang2018infrastructure}. However, a critical challenge in modeling such parameters is that the vehicle's kinematic/dynamic information can be lost due to non-linearity and complexity of dynamics, particularly when a vehicle is traversing under a high speed, a large curvature, or a unique condition \citep{yao2020control}. A recent empirical study also showed that control sensitivity factors (i.e., control gains that regulate the behavior) could vary depending on speed and headway settings, thus signifying highly nonlinear control mechanisms \citep{shi2021empirical}. Thus, the parameters of the GLVD equation are highly stochastic and correlated with the traffic state and even geometric and environmental conditions.

Given the profound impact on CF control performance and stability, the stochasticity of GLVD vehicle dynamics parameters should be addressed in a real-time setting. The principal idea here is that we cannot fully account for all exogenous/endogenous uncertainties that might affect the vehicle's performance while traversing the physical world simply because we cannot ascertain the nature of these uncertainties. Thus, it is beneficial to allow the performance of an AV in real-time to speak for itself. We do so by utilizing the sensor data by an AV (speed, acceleration, jerk, position, etc.) and estimating the real-time value of the GLVD parameters to see how much uncertainty is occurring in our system.

Note that parameter estimation techniques have been employed in some control systems to address stochasticity in control parameters and improve control accuracy. Some notable algorithms include the Kalman filter, least-squares error estimate, parameter identification, reinforcement-learning methods, and neural predictive networks. \cite{yao2020control} reviews some applications of these algorithms and presents the respective disadvantages of each algorithm. For instance, these algorithms often fail to scale in an online setting, can require a large amount of data, or yield large errors when dynamics are non-linear. Nonetheless, parameter estimation techniques could provide effective means to address parameter uncertainty. Yet, their usage in vehicular dynamics, particularly GLVD parameters, is yet to be used. Additionally, and on a more critical note, the fundamental shortcoming of the application of such methods is the inability of users/designers to take strategic actions on the performance of the AV in real-time if anomalies (e.g., performance loss due to large uncertainties ) are detected. The benefit of strategic-actions, is that only through it we can connect the individual performance of an AV, to the overall performance of the traffic system containing the AV. This particular challenge is what this work aims to tackle. To the extent of the author's knowledge, there are not yet any existing methodologies or applications in the literature that tackles this challenge.

We conjecture that addressing exogenous and endogenous uncertainties in CF control should be done both at the modeling level (e.g., robust control design) and the decision-making level (e.g., dynamic updating of control parameters). This paper focuses on the latter and aims to develop a methodology to estimate uncertainties in vehicle dynamics, as modeled in GLVD, in real-time and monitor the performance of CF control to enable strategic decisions if needed. Specifically, the methodology comprises three elements: (i) Bayesian inference of parameter uncertainties in GLVD that can be dynamically updated in response to different driving scenarios and capture unobserved stochasticity, (ii) dynamic monitoring of CF performance such as stability, and (iii) strategic adjustment of parameters to improve the CF control performance if an anomaly is detected. Note that Bayesian inference is usually computationally demanding, which renders online applications impractical. To overcome this issue, we utilize Stochastic Gradient Langevin Dynamics (SGLD) to formulate a stochastic optimization problem that estimates uncertainties in GLVD parameters in real-time and continuously update the estimates on the fly with new sensor measurements. Building on this, we develop a monitoring methodology that continuously assesses the performance of the CF controller, specifically local and string stability requirements. This integrated framework allows for strategic adjustment of controller parameters to induce desired performance in a more adaptive manner. Therefore, the main novelty of this paper lies in adaptive AV CF control enabled by the integrated framework of real-time uncertainty quantification and dynamic monitoring of the CF controller performance. 

The rest of the paper is organized as follows: Section 2 presents a background analysis of the impacts of parameter uncertainties of vehicular dynamics on AV CF stability. Section 3 provides the details of the mathematical formulation for our real-time uncertainty estimation model. Then Section 4 describes the parameter monitoring schemes and how they are combined with strategic parameter adjustment to preserve the desired performance of the controller. Finally, discussion and concluding remarks are provided in Section 5.

%%----------------------------------------------------------
\section{Background}\label{S:2}
This section introduces the basis of linear CF controller - which will be used here to showcase our modeling and applicability in different scenarios - and highlights the potential impact of vehicular dynamics on performance. Specifically, we show how the GLVD equation can be incorporated into a typical linear controller and the underlying impacts on local and string stability. 

\subsection{Linear Controller Background}
Linear controllers are widely adopted control algorithms, primarily due to their analytical properties and stability guarantees. They have rich theoretical and methodological literature and have recently shown great promise in real-life applications, specifically on ACC/CACC systems \citep{shladover2015cooperative,li2021car,gunter2020commercially,milanes2014modeling,morbidi2013decentralized}. While various linear control systems exist in the literature, their underlying control strategy is relatively consistent. In this paper, we focus mainly on robust linear controllers that incorporate vehicular dynamics. Specifically, we focus on the state-of-the-art controller developed by \cite{zhou2019robust} for demonstration purposes. It provides mathematical formulations of local/string stability requirements while incorporating vehicular dynamics and communication delays. Note that the overall methodology developed in the present paper is primarily data-driven (as will be shown in Sec.\ref{S:3}) and thus can be adapted to different control paradigms (e.g., MPC). 

The controller developed by \cite{zhou2019robust} follows a  hierarchical design whereby the upper-level controller regulates the CF behavior based on the widely adopted constant time gap policy (CTP). Accordingly, the state space formulation is defined as $\bm x(t) = [\Delta s(t), \Delta v(t), a(t)]^T$, where $\Delta s(t)$ is the deviation from target spacing defined by a constant time gap $\tau^*$, $\Delta v(t)$ is the speed difference with the leading vehicle, and $a(t)$ is the actual acceleration of the vehicle. 

Notably, this design assumes that acceleration is not perfectly implemented due to various uncertainties in vehicular dynamics (e.g., vehicle condition, gear position, aerodynamic drag, road gradient, etc.). Accordingly, a lower-level controller is added, which adopts the GLVD equation to incorporate such uncertainties. The formulation of vehicle dynamics in GLVD is shown in Eq. \ref{eq:glvd} \citep{yi2001vehicle,wang2018infrastructure}. 

\begin{equation}
\label{eq:glvd}
\dot{a}(t) = \frac{-1}{T_L}a(t) + \frac{K_L}{T_L}u(t) + \epsilon(t)
\end{equation}\\
where $\dot{a}(t)$ is the vehicle's jerk, $a(t)$ is the actual acceleration, $u(t)$ is the demanded acceleration by the controller at time $t$. $T_L$ is the actuation time lag, and $K_L$ is the ratio of the demanded acceleration that can be realized. $u(t)$ is determined through the feedback control law as $u(t) = k x(t)$, where $k = [k_s, k_v, k_a]$ is a vector of feedback control gains that regulates $\Delta s(t)$, $\Delta v(t)$, and $a(t)$, respectively. $\epsilon(t)$ is an additive error term. (Note that a feedforward gain parameter that considers uncertain communication delays in the presence of vehicle-to-vehicle or vehicle-to-infrastructure communication is added in \cite{zhou2019robust}; however, for conciseness, we will not consider such uncertainties.)  

One can clearly notice how parameters $T_L$ and $K_L$ influence the actual acceleration of the vehicle as compared to the demanded acceleration by the controller. Previous literature has also highlighted the relation of parameters $T_L$ and $K_L$ to uncertainties in vehicular dynamics \citep{li2018robust,trudgen2015robust,gao2016robust}. Ideally, if no external disturbances affect the controller, $a(t) = u(t)$, and thus $T_L=0$ and $K_L=1$. In reality, however, $T_L$ and $K_L$ are highly stochastic parameters and carry significant impact on the local/string stability of the vehicle as highlighted below.

\subsection{Impact of Vehicular Dynamics on Local and String Stability}

Local stability (dissipation of disturbances over time) and string stability (attenuation of disturbances over a vehicular string) are two critical attributes in developing safe and effective automated traffic. In \cite{zhou2019robust}, sufficient and necessary local and string stability conditions are mathematically derived as shown below (readers are referred to the cited paper for the proof and details, we only show here some formulation for the sake of illustration):

\begin{itemize}
    \item \textbf{Sufficient and necessary conditions for local stability}:
    \begin{equation}
    \begin{cases}
        1-K_L^{u}k_a &> 0 \\
        k_s\tau^* + k_v &> 0 \\
        k_s &> 0 \\
        \Big(\frac{1}{K_L^l} -k_a\Big)\Big(k_s*\tau^* + k_v\Big) &> \frac{T_L^u}{K_L^l}k_s\\
        \Big(\frac{1}{K_L^u} -k_a\Big)\Big(k_s*\tau^* + k_v\Big) &> \frac{T_L^l}{K_L^u}k_s 
    \end{cases}
    \end{equation}
    
        \item \textbf{Conditions for string stability}:
    
    \begin{equation}
    \begin{cases}
        \Big(K_L^uk_a -1\Big)^2 - 2T_L^uK_L^u\Big(\tau^*k_s + k_v\Big) &> 0\\
        \Big(K_L^lk_a -1\Big)^2 - 2T_L^uK_L^u\Big(\tau^*k_s + k_v\Big) &> 0\\
        K_L^l\Big[2k_sk_a + (\tau^*k_s + k_v)^2 - k_v^2\Big] -2k_s &> 0\\
    \end{cases}
    \end{equation}
\end{itemize}

In a typical controller, the control gains $k=[k_s, k_v, k_a]$, time gap setting $\tau^*$, and parameters $T_L$ and $K_L$ are preset for each vehicle. However, parameters $T_L$ and $K_L$ are stochastic in nature, and there is no guarantee that they remain consistent in real-life conditions. For instance, the recent empirical analysis on commercial ACC-enabled vehicles by \cite{gunter2020commercially} estimated systematic delays in system sensors to be $0.1-1$ seconds. Note that these delays are related to second-order lags (i.e., pertaining to speed and location), whereas $T_L$ and $K_L$ are related to third-order lags. While the estimates of second-order lags are not strictly equivalent to $T_L$ or $K_L$, they suggest the significant impact of these parameters and variation in experienced delays by real-life vehicles. \citep{wang2018delay} shows the impact of different actuation lag values on the acceleration profile of the vehicle controller and its overall stability. 

\begin{figure}[!htb]
\centering
    \begin{subfigure}[b]{0.45\textwidth}
      \centering
      \includegraphics[width=0.7\textwidth]{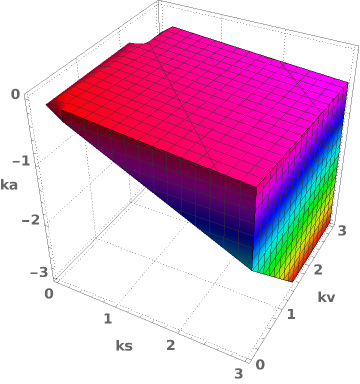}
      \caption{$[T_L^l,T_L^u]=[0,0.1]$}
    \end{subfigure}
    \hfill
    \begin{subfigure}[b]{0.45\textwidth}
      \centering
      \includegraphics[width=0.7\textwidth]{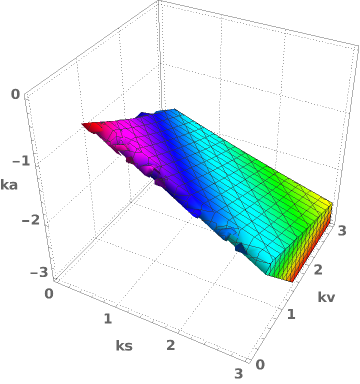}
      \caption{$[T_L^l,T_L^u] = [0,1]$}
    \end{subfigure}
\caption{Stability regions for different $T_L$ settings}
\label{fig:timpact}
\end{figure}

\begin{figure}[!htb]
\centering
    \begin{subfigure}[b]{0.45\textwidth}
      \centering
      \includegraphics[width=0.7\textwidth]{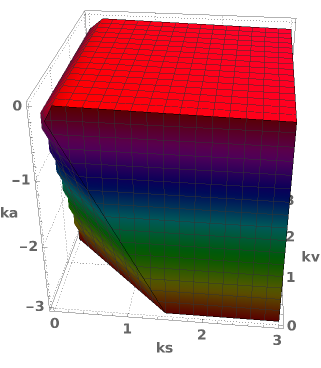}
      \caption{$\tau^* = 2$}
    \end{subfigure}
    \hfill
    \begin{subfigure}[b]{0.45\textwidth}
      \centering
      \includegraphics[width=0.7\textwidth]{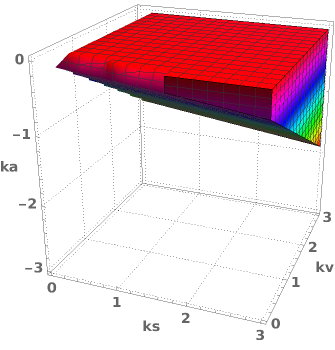}
      \caption{$\tau^* = 0.5$}
    \end{subfigure}
\caption{Stability regions for different $\tau^*$ settings}
\label{fig:taoimpact}
\end{figure}

In the formulation above $T_L$ and $K_L$ are assumed to be random but bounded: i.e., $0<T_L^l\leq T_L\leq T_L^u$ and $0<K_L^l\leq K_L\leq K_L^u$. In essence, the vehicle is stable given that $T_L$ and $K_L$ remain within the bounded regions. If these parameters vary beyond these limits due to uncertain conditions in real life, stability will be impacted. Further, even within the desired bounds, variability in these parameters may hamper the effectiveness of resolving disturbances. To visualize the extent of the impact, we simulate the stability regions in different scenarios. Specifically, we highlight the impact of the actuation lag $T_L$ and time gap setting $\tau^*$ as a function of control gain settings. Note that these figures are only intended to showcase general patterns rather than deep quantitative analysis. In a previous work by the authors, we showcase how stability regions are impacted by controller gains \citep{kontar2021multi}, and \citep{zhou2019robust} provides deeper insights into theoretical stability region. Fig. \ref{fig:timpact} shows that when $T_L$ increases, the stability regions shrink significantly. This is expected as with a large actuation lag, the actual acceleration of the vehicle is far off the demanded acceleration by the controller, which leads to undesired performance. This further suggests that the misspecification of this parameter can yield unstable behavior. Notably, the time gap setting plays another significant role on the stability region, as shown in Fig. \ref{fig:taoimpact}. In essence, $\tau^*$ affects the response duration of the controller and how the controller responds to disturbances. An increase in $\tau^*$, would diminish the impact of disturbances on the vehicle and thus increase its stability region. Additionally, $K_L$ was seen to exhibit an impact on the stability region, yet to a less extent than $T_L$. The above analysis emphasizes the need for dynamic actions to adjust controller parameters to remain within the stability regions if $T_L$ and $K_L$ deviate from the desired settings. The strategic adjustment of the control will be discussed in further detail in Sec. 4. However, the principal idea is to adjust the control gain setting ($k$) or time gap parameter ($\tau^*$) in ways to regain local and string stability.

The stochastic nature of parameters $T_L$ and $K_L$ embedded in their relation to real-time conditions, along with their impact on the performance of the CF controller presents a challenging yet vital factor in achieving the desired CF behavior. In the following sections, we present a modeling framework that can estimate parameters $T_L$ and $K_L$ in real time, gauge the uncertainty in vehicular dynamics, and enable strategic decisions in real-time to amend any anomalies in performance.

%---------------------------------------------------------------
\section{Real-time Uncertainty Estimation: Model Formulation}\label{S:3}
In this section, we present the formulation of a real-time estimation model for the stochastic parameters $T_L$ and $K_L$. The proposed model adopts a Bayesian approach that utilizes real-time sensor measurement data to estimate dynamically and update the uncertainties related to vehicular dynamics in the GLVD equation. 

\subsection{Model Basics}

A linear controller incorporates vehicular dynamics using the GLVD equation shown in Eq. \ref{eq:glvd} by regulating the demanded acceleration ($u(t)$) based on $T_L$ and $K_L$. Thus, for dynamic monitoring of vehicle performance, we propose an online estimation method for $T_L$ and $K_L$ based on Bayesian inference to quantify uncertainties in these parameters continuously. Information necessary for this process comprises the vehicle's acceleration and jerk and demanded acceleration from the controller, all of which are attainable by the vehicle through sensor measurements (for acceleration and jerk) and its control system (for demanded acceleration). 

More specifically, consider a stream of data where dataset $D_t$ is observed at each time instance $t \in \{1,.., T\}$. Here the horizon $T$ need not be finite.  Now, let $N_t$ be the number of observations obtained at $t$ and let $D_t=\{\bm{U}_t, \bm{\dot{A}}_t\}$ where $\bm{U} \in \mathbb{R}^{N_t \times 2}$ denotes the input matrix from $a(t)$ and $u(t)$ and $\bm{\dot{A}} \in \mathbb{R}^{N_t \times 1}$ denotes the output vector corresponding to $\dot a(t)$. We also define $D_{\tilde{t}}=\{D_1, ..., D_t\}$ to denote all data observed up to $t$.

Our overarching goal is to: (i) estimate $K^t_L$ and $T^t_L$ at each $t\in \{1,..., T\}$, (ii) quantify uncertainties in $K^t_L$ and $T^t_L$, and (iii) continuously update $K^t_L$ and $T^t_L$ on the fly when $D_{t+1}$ is attained. To this end, we take a Bayesian approach to estimate the posterior $p(K^t_L, T^t_L|D_{\tilde{t}})$. This allows for uncertainty quantification yet induces two fundamental challenges: (i) $\dot a(t)$ is non-linear with respect to $K^t_L$ and $T^t_L$; hence conjugate prior schemes to obtain a closed-form posterior distribution are not a viable option, and (ii) $p(K^t_L, T^t_L|D_{\tilde{t}})$ needs to be attained and updated in real-time, often within few seconds (the exact update frequency can be a user choice; say every 5 seconds we collect data and then perform the update). The latter renders sampling schemes such as Markov Chain Monte Carlo (MCMC) \citep{brooks1998markov} unviable options due to their computational complexity, as they require the entire sample space $D_{t}$ at each iteration to do the posterior computations. 

To address these challenges, we develop an empirical method that collects data samples belonging to the posterior distribution and then estimates the distribution through the collected samples. In Bayesian statistics, one common method that employs this principle is the Maximum a Posterior (MAP) estimate, which follows Bayes' theorem and gives the mode of the posterior distribution (i.e., the most frequent data point) through empirical data (i.e., samples). Note that the Maximum Likelihood Estimator (MLE) is similar to MAP in its empirical intuition. However, with its Bayesian root, MAP can lead to the estimation of the posterior distribution, not only the likelihood as in MLE. Note further that MAP itself gives a point estimate, not a posterior distribution. This approach will be further extended shortly to generate a posterior distribution.

Let us first explore the formulation of the MAP for our application. Following Bayes' theorem, $P(K^t_L, T^t_L|D_{\tilde{t}}) \propto P(K^t_L, T^t_L) \times P(D_{\tilde{t}}|K^t_L, T^t_L)$. Here $P(D_{\tilde{t}}|K^t_L, T^t_L) = \prod_i^{N_{\tilde{t}}} P(\dot{a}_i|a_i, u_i, K^t_L, T^t_L)$ where $N_{\tilde{t}}$ is the cardinality of $D_{\tilde{t}}$ and $a_i$, $u_i$, $\dot{a}_i$ are the observations within $D_{\tilde{t}}=\{\bm{U}_{\tilde{t}}, \bm{\dot{A}}_{\tilde{t}}\}$. As a result, the MAP can be obtained as:

\begin{equation}\label{eq:map}
    \underset{K^t_L, T^t_L}{\textrm{min}} \quad -\left(\log P(K^t_L, T^t_L) +  \sum_i^{N_{\tilde{t}}} \log P(\dot{a}_i|a_i, u_i, K^t_L, T^t_L)\right)  
\end{equation}
where $\log P(\dot{a}_i|a_i, u_i, K^t_L, T^t_L)$ is written as the log Gaussian likelihood, $-\frac{1}{2}\log\sigma^2 - \frac{1}{2\sigma^2}\left(\dot{a}_i-\dot{a}(t)\right)^2$. Using the log Gaussian likelihood comes with the assumption that $\epsilon(t) \sim \mathcal{N}(0, \sigma^2)$, where $\epsilon(t)$ is the additive error/noise parameter in Eq. \ref{eq:glvd}. 

As mentioned, solving the MAP (Eq. \ref{eq:map}) for the mode of the posterior distribution is not the end goal. We wish to estimate the entire posterior distribution, which means collecting more samples and then estimating this distribution empirically. Accordingly, another challenge is what solving algorithm for Eq. \ref{eq:map} allows for the extraction of samples that belong to the posterior distribution. In the following subsection, we describe this in more detail.

\subsection{The Stochastic Gradient Langevin Dynamics}

Based on the discussion above, we now concern ourselves with solving algorithms that: (i) have fast computation for real-time estimation, (ii) allow for uncertainty propagation, and (iii) allow for estimation of the posterior distribution. Typically, the MAP shown in Eq. \ref{eq:map} can be solved using the Stochastic Gradient Descent (SGD). The idea behind SGD is that we take iterative steps in the direction of steepest descent to reach a local minimum. Formally we express this for any parameter $\theta$ and a dataset $D$, SGD will find the gradient of Eq. \ref{eq:map} using a minibatch of the data ($\tilde{D} \subset D$) of size $\tilde{N}$ as

\begin{equation}\label{eq:sgd}
    \nabla\theta^{(t)} = \frac{\eta_t}{2}\Bigg( \nabla\log P(\theta^{(t)}) + \frac{\tilde{N}}{n}\sum_{i=1}^n \nabla \log P(D^{(t,i)} | \theta^{(t)}) \Bigg)
\end{equation}
where $\eta_t$ is a sequence of step sizes of the gradient descent at iteration $t$. The parameter updates are obtained by a simple gradient decent step $\theta^{(t+1)} = \theta^{(t)} - \eta_t\nabla\theta^{(t)}$. Through iterating the SGD in Eq. \ref{eq:sgd}, parameters can converge to a critical point on the condition of decreasing step sizes $\sum^{\infty}_{t=1}\eta_t = \infty$ and $\sum^{\infty}_{t=1}{\eta_t^2} < \infty$. 

SGD has been widely adopted in various optimization methods due to its simplistic form and fast computation. However, its applicability in our envisioned model is limited since it does not capture parameter uncertainty nor does it allow us to estimate the posterior distribution. Implementing SGD on Eq. \ref{eq:map} would simply result in the solution for the mode of the posterior distribution, not the distribution. This issue is remedied by adopting the principle of Stochastic Gradient Langevin Dynamics (SGLD) to solve the MAP in Eq. \ref{eq:map}

\subsubsection{Posterior Sampling}

The mechanics of SGLD \citep{welling2011bayesian} are similar to those of SGD. However, in SGLD, a Guassian noise is injected into Eq. \ref{eq:sgd} that depends on the step size $\eta_t \sim N(0,\eta_t\bm I)$. The Gaussian noise prevents parameter updates to collapse to just the MAP solution, but allows convergence to samples from the true posterior distribution. This approach makes SGLD efficient with large datasets while providing a much-needed Bayesian interpretation to parameter uncertainty. Specifically, for our real-time estimation scheme at each iteration $t$, a mini-batch $\{D^{(\tilde{t},i)}\}_{i=1,2,\cdots,N_{\tilde{t}}}$ of size $N_{\tilde{t}}$ is used to update the SGLD parameters according to:

\begin{align}\label{eq:sgld}
    &\nabla(K^t_L,T^t_L) = \frac{\eta_t}{2}
    \Bigg( \nabla\log P(K_L^t,T_L^t) + 
    \frac{N_{\tilde{t}}{n}}
    \sum_{i=1}^{n}\nabla\log P(\dot{a}_i | a_i, u_i, K_L^t, T_L^t)
    \Bigg) 
    + \epsilon_t \\
    &\epsilon_t \sim \mathcal{N}(0,\eta_t\bm I)
\end{align}

\subsubsection{Intuition}
In SGLD, we add Gaussian noise to each step, $\eta_t$. This induces noise along with the learning rate that will then decay with each iteration. The idea is that SGLD allows the optimization to locate the mode (i.e., MAP); however, it never actually collapses to it due to the added noise. Moreover, it also does not allow the solution to leave the locality of the mode because the learning rate is decaying. Eventually, SGLD iterations will be randomly walking around the MAP collecting samples (which are theoretically proven to belong to the true posterior distribution; see proof in \citep{welling2011bayesian}. Accordingly, there are two phases for the SGLD: (i) optimization in which iterations locate the mode, and (ii) sampling in which iterations walk around the mode. Ultimately, this allows us to collect parameter values during the sampling phase and thus enabling Bayesian learning to estimate the posterior distribution. Further, another nice aspect about SGLD is that we can do all this through an optimization framework with mini-batch data (i.e., subset of data). For this reason, SGLD is very fast, computationally effective, and overcomes curse of dimensionality.  

\begin{figure}[!htb]
\centering
\includegraphics[width=0.6\linewidth]{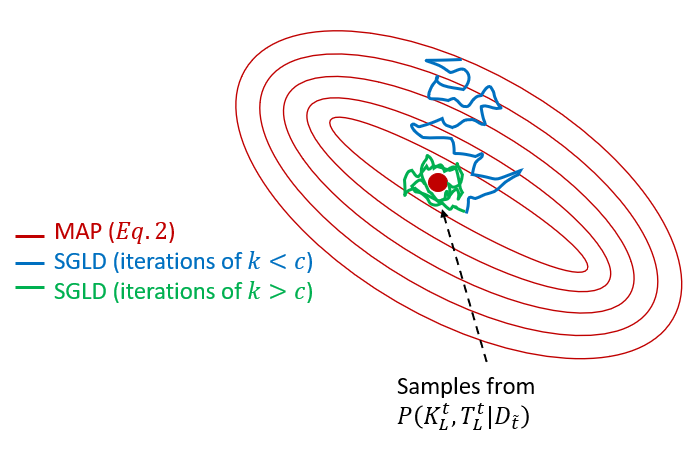}
\caption{SGLD algorithm depiction}
\label{fig:algorithm}
\end{figure}

We depict this  in Fig. \ref{fig:algorithm}. Note that the figure is not drawn to scale, and the region of posterior sampling (green color) is in fact relatively large. We further summarize the developed model in Algorithm \ref{alg:algorithm1}. Note that here $c$ - upon which the parameter starts sampling - is relative to the decay of learning rate and step sizes $\eta_t$ and $\epsilon_t$.

\begin{algorithm}[t]
\SetAlgoLined
 Input: ${\bm\theta}^{(0)} = [K^0_L,T^0_L]\in \mathbb{R}^2$, initial step size $\eta_1>0$.\\
\For{$k=1,2,\dotsc,K$}
  {Randomly sample a subset of the minibatch $\{D^{(k,i)}_{i=1,...,N_{k}}\}$ of size $N_{k}$\;
  Compute the stochastic gradient $\nabla(K^k_L,T^k_L)$\;
  %Select a stepsize $\alpha_k>0$\;
  $\eta_k\leftarrow \frac{\eta_1}{k}$\;
  $\bm\theta^{(k)}\leftarrow\bm\theta^{(k-1)} - \eta_k\nabla(K^k_L,T^k_L)$\;}
\For{$k>c$}
{Collect output $\bm\theta^{(k)}\sim P(K^t_L, T^t_L|D_{\tilde{t}})$}
\caption{SGLD Parameter Estimation}
\label{alg:algorithm1}
\end{algorithm}%\vspace{-2mm}

\subsubsection{Update with New Data}
Now that we have collected samples from the posterior distribution $P(K^t_L, T^t_L|D_{\tilde{t}})$ through SGLD, we can empirically estimate the parameters $T_L$ and $K_L$. More precisely, we can define the empirical cumulative distribution function ${F}(\tilde{N_t})$ as:

\begin{equation}
   {F}(\tilde{N_t}) = \frac{1}{\tilde{N}_t}\sum_{i=1}^{\tilde{N}_t}{\theta^{(i)}} 
\end{equation}

Our approach is a dynamic and iterative one. Specifically, we are interested in continuously updating the posterior distribution on the fly once a new set of data is obtained. Here we describe how to update with new data. 

Following Bayes' rule, the update process would use the estimated posterior distribution at time $t-1$ as a prior to estimate the posterior at time $t$. While this can be achieved by simply using posterior as prior, we seek to integrate the fundamentals of regularization to achieve a more accurate updating scheme. Specifically, we tweak the prior distribution, $P_{prior}$, in a way that penalizes large deviations from the previous estimated posterior distribution when estimating with new data. 

More specifically, we will define our prior distribution as:

\begin{equation}
    \mathbb{P}^{(t)}_{prior} \sim \mathcal{N}\left(\left[\bar{K}_L^{(t-1)},\bar{T}_L^{(t-1)}\right]^{\top},\lambda\bm I \right)
\end{equation}\\
where $\lambda$ is a regularization term that is tunable. Note that the regularization happens only to the variance of the prior $\left(\lambda \bm I \right)$, but the mean remains the same. The intuition behind this regularization is to allow the posterior estimation to focus better on new data while preserving previously acquired accuracy.

With the adoption of $\mathbb{P}^{(t)}_{prior}$, we can reach an equivalent formulation of our MAP expressed in Eq. \ref{eq:map}, as below:

\begin{equation}
   \underset{K^t_L, T^t_L}{\textrm{min}} \quad \left( 
    \sum_{i=1}^{N_{\tilde{t}}}  \left[\dot{a}_i - \dot{a}_i\left(\left[K_L^{(t)},T_L^{(t)}\right]\right)
    \right]^2
    - \frac{\sigma^2}{\lambda} \left\lVert\, \left[\bar{K}_L^{(t)},\bar{T}_L^{(t)}\right]^{\top} - \left[\bar{K}_L^{(t-1)},\bar{T}_L^{(t-1)}\right]^{\top} \right\rVert_2^2
    \right)
\end{equation}\\

\textbf{Some Notes on the developed model:}\\
\noindent \textbf{\emph{1. On-time horizon $T$:}} The time horizon is a matter of choice. Within the specified time horizon, we are collecting and storing a batch of data that is acquired in real-time (through sensor measurements, controller data, etc...). At the end of the horizon, we use the collected data and perform estimation in an iterative fashion. We note that our update scheme (Sec. 3.2.3.) makes sure the complexity will not explode over time, as we perform dropping of previous data and use on the prior distribution. Thus, users are able to tune the time horizon to suit their needs, albeit not myopic. Since ultimately, SGLD uses a mini-batch to perform iterations, the computation time is significantly fast and will not become a burden in real-time operation. \\
\noindent \textbf{\emph{2. On generalizability of application:}} While in this paper we focus on certain types of system uncertainties, the approach of real-time Bayesian estimation remains generic. With the increase in real-time data availability for AVs, the proposed technique can be further expanded to tackle different types of uncertainties.\\
\noindent \textbf{\emph{3. Measurement Errors:}} Since our model relies on sensor measurement, errors in these measurements might impact the overall performance. Although we have an error term to account for that in the formulation. 

\subsection{Model Performance}
To demonstrate the proposed estimation method, we run a series of controlled simulation experiments, where we use a lead vehicle trajectory extracted from the NGSIM data to present realistic driving situations and then simulate an AV trajectory using the controller discussed in Sec. 2. We run multiple trials, where at each trial we change the value of $T_L$ or $K_L$ in the controller while controlling for other parameters (i.e., gains). We first assume a non-informative prior distribution of $T_L$ and $K_L$ and then run Algorithm 1 for 3 iterations. Each iteration represents 5-second intervals, with the data frequency of 0.01 secs, meaning that at each iteration 500 data points are used for estimation. After 3 iterations, the estimated distributions represent an uninformative-prior which is used to initiate the iterations afterward. Then, we run our model for 7 new iterations and extract the final posterior distributions and obtain the final estimates of these parameters. 

\begin{figure}[!htb]
\centering
    \begin{subfigure}[b]{0.45\textwidth}
      \centering
      \includegraphics[width=\textwidth]{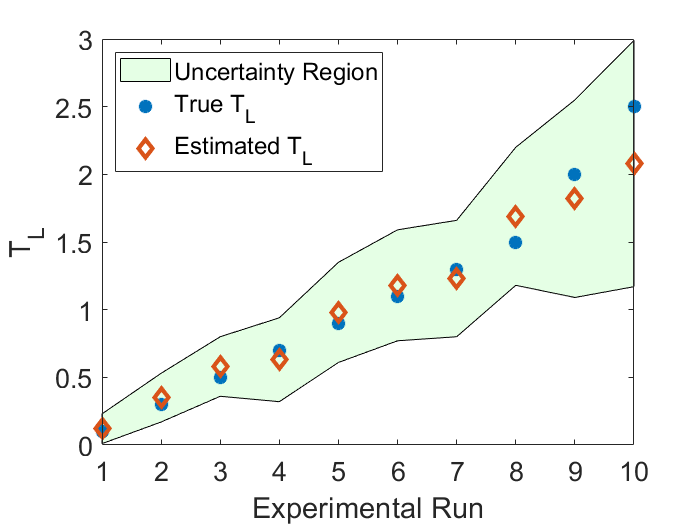}
      \caption{$T_L$ estimation}
      \label{fig:tlestimation}
    \end{subfigure}
    \hfill
    \begin{subfigure}[b]{0.45\textwidth}
      \centering
      \includegraphics[width=\textwidth]{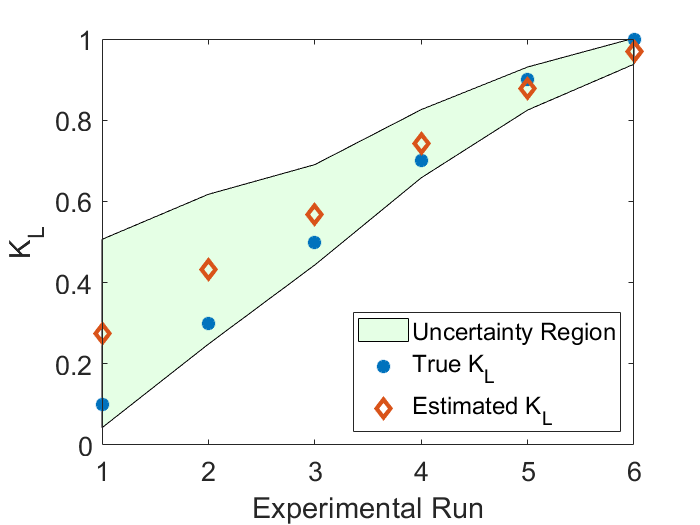}
      \caption{$K_L$ estimation}
      \label{fig:klestimation}
    \end{subfigure}
\caption{Parameter estimation through SGLD}
\label{fig:sgldexperiments}
\end{figure}

Fig. \ref{fig:sgldexperiments} summarizes the results of the experiments. We notice that the model can track the parameters $K_L$ and $T_L$ while estimating an uncertainty bound around them. Notably, the error in estimation increases when $T_L$ increases (i.e., deviates away from 0), and similarly error increases when $K_L$ decreases (i.e., deviates away from 1). In cases of high (low) $T_L$ ($K_L$), the controller becomes more unstable and less effective in dissipating disturbances, which leads to more noise in the controller output (e.g., acceleration). However, the estimation of uncertainty is particularly helpful in such cases, as the true value still lies within the upper and lower bounds of the estimation. In the following section, we discuss what strategic actions can be taken to address such cases to maintain desirable performance. 

\section{Real-Time Strategies: Monitoring and Adjusting Control Parameters}\label{S:4}
Up till now, we have outputted estimated distributions of $T_L$ and $K_L$ at every estimation period. Given these estimates, we are now interested in profiling and monitoring these parameters through time and take strategic decisions if any discrepancies in performance are expected. In the following section, we showcase how multiple real-time strategies can be built upon the above estimation to maintain desired CF performance under uncertainty. 

%\subsection{An Overview of the Strategic Approach and its Integration into the Control System}

\begin{figure}[!htb]
\centering
\includegraphics[width=0.8\linewidth]{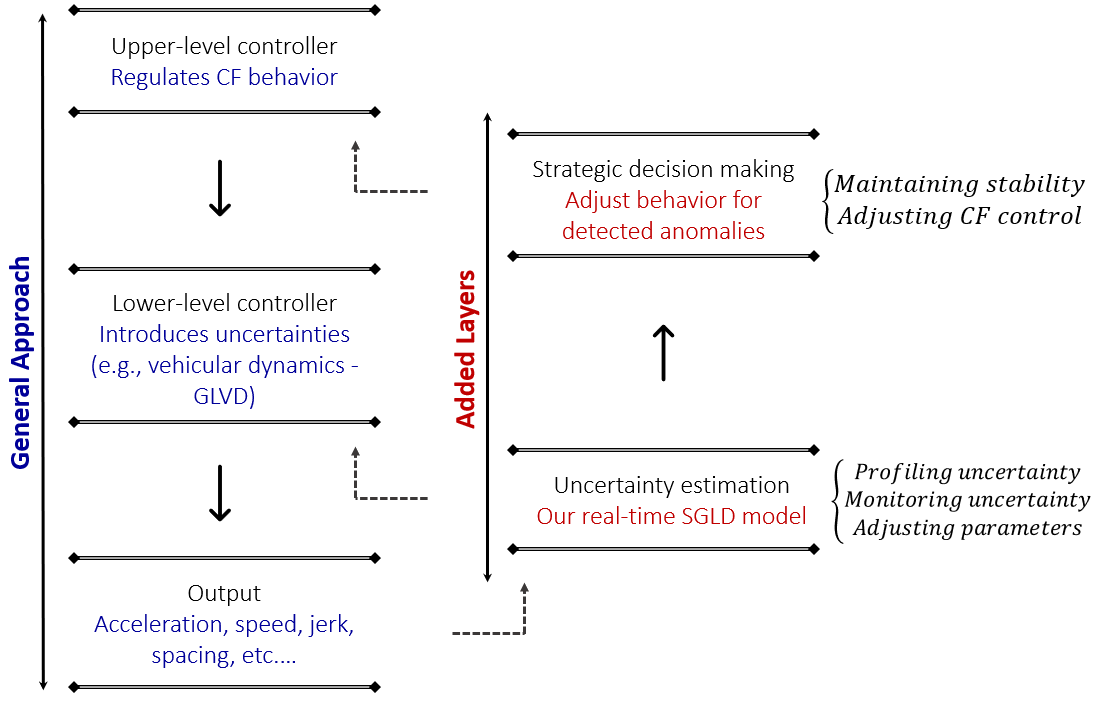}
\caption{An illustration of the strategic approach}
\label{fig:strat}
\end{figure}

The developed real-time estimation model can be adopted by itself in controllers to make them more robust against system uncertainties, for example, by updating the bounds of the parameters, $T_L$ and $K_L$. However, we believe that this is only one layer of protection against real-world problems, and more proactive dynamic control adjustment can be done to realize more adaptive and effective control. To enable dynamic adjustment, we focus on (i) monitoring the deviations of $T_L$ and $K_L$ from acceptable values, (ii) continuously examining whether the CF controller is within local/string stability regions, and (iii) determining the best action for control adjustment (i.e., what parameter to adjust). This allows us to gauge the performance of the controller in real-time and adjust its behavior (i.e., control parameters) to recover the desired performance. Fig. \ref{fig:strat} depicts the general approach of our decision-making scheme.

\newpage

%\subsection{Application Analysis}
Here we present three different strategies for control adjustment. Strategy 1 pertains to the estimation and adjustment of $T_L$ and $K_L$ in the lower-level controller. Strategy 2 involves adjustment of the time gap in the upper-level controller. Strategy 3 involves adjustment of control gains in the upper-level controller. Strategy 1 should be adopted at a minimum. Strategy 2 and/or 3 should be adopted, in addition to strategy 1, when a more drastic action is needed. These strategies are demonstrated with simulation experiments. 

\textbf{Remark}: \emph{In Sec. 4, our aim is to present and lay a foundation of possible strategies that can be done through the methodology introduced in this paper. However, the detailed design of the strategy can be left to the users/practitioners/designers to do as they deem fit to their application. We further explain in Sec. \ref{S:4.5} how users can personalize the strategic-decisions. }

\subsubsection{Simulation Setup}
\begin{figure}[!htb]
\centering
    \begin{subfigure}[b]{0.45\textwidth}
      \centering
      \includegraphics[width=\textwidth]{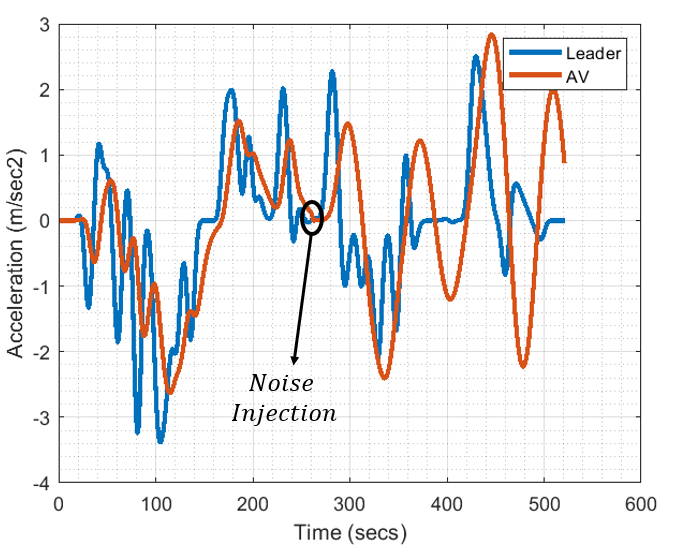}
      \caption{Acceleration profile of leader and follower showing instance of noise injection}
      \label{fig:openacc}
    \end{subfigure}
    \hfill
    \begin{subfigure}[b]{0.45\textwidth}
      \centering
      \includegraphics[width=\textwidth]{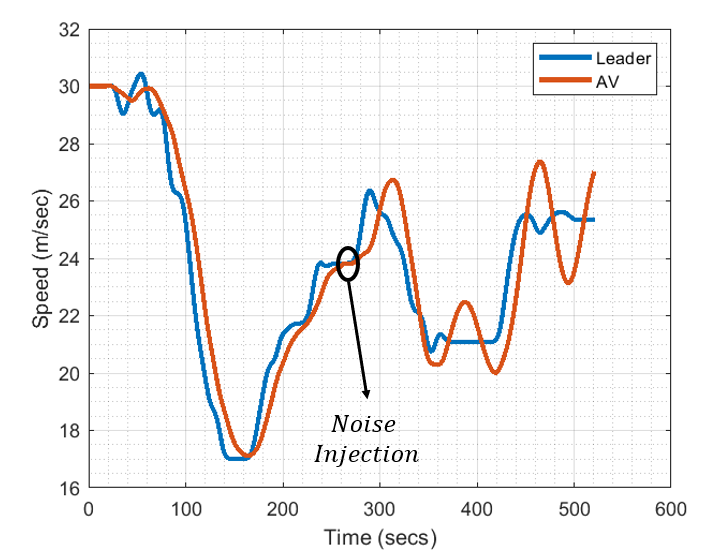}
      \caption{Speed profile of leader of leader and follower showing instance of noise injection}
      \label{fig:openspeed}
    \end{subfigure}
\caption{Simulation setup}
\label{fig:leader}
\end{figure}

In order to make our simulation experiment as realistic as possible, we adopt the OpenACC dataset \citep{makridis2021openacc} that represents real-life driving scenarios of commercial ACC vehicles. Specifically, we choose a trajectory profile from the OpenACC dataset and then construct its acceleration profile. Note that we perform Gaussian smoothing on the acceleration profile. Taking this as a leader trajectory we then simulate a response by an AV, according to the controller in Sec. 2. Precisely, our parameter settings for the controller are presented in Table \ref{tab:simparam}. To simulate stochasticity impacted from vehicular dynamics, we inject noise into the AV acceleration profile, as seen in Fig. \ref{fig:openacc}. The noise injected is designed to replicate a vehicular dynamic uncertainties (relative to GLVD) for $T_L = 1.5$ and $K_L = 0.5$. While these values are not rigorous estimates from the data as the controllers are unknown, inspection of the data reveals that they are in the possible range. We also note that the selected values might be more on the extreme spectrum, but this helps us get better insights into our model performance and different potential strategies we can implement.

\begin{table}[!htb]
\centering
\caption{Default parameter settings for controller}
    \begin{tabular}{ll}
    \hline
    Parameter & Value\\
    
    \hline
        $T_{L}$ & 0.3 secs\\
        $K_{L}$ & 1 \\ 
        $t_s$ & 0.01 secs \\
        $\tau^*$ & 1 sec \\
        $\delta^*$ & 5 m\\
        $K$ & $[1.5, 1.5, -0.8]$\\
    \hline
\end{tabular}
\label{tab:simparam}
\end{table}

\newpage

\subsubsection{Strategy 1: Parameter Monitoring \& Adjustment in Lower-Level Control}

This strategy entails profiling and monitoring parameters $T_L$ and $K_L$ based on the real-time estimation scheme developed in Sec. 3, and adjusting their values when an anomaly is detected. Specifically, we can run Algorithm 1 and estimate these parameters with each given dataset from the sensors. This allows us to detect when these parameters are out of desired values. For instance, in the robust controller discussed in Sec. 2, the $T_L$ and $K_L$ are assumed to be bounded, with preset bounds. These bounds are typically tight, with \cite{zhou2019robust} specifying the bounds of $[0.1,0.4]$ and $[0.7,1]$ for $T_L$ and $K_L$, respectively. 

\begin{figure}[!htb]
\centering
    \begin{subfigure}[b]{0.45\textwidth}
      \centering
      \includegraphics[width=\textwidth]{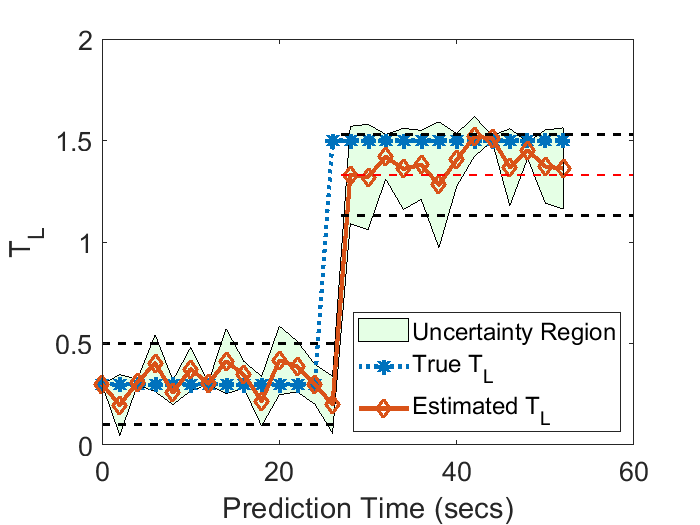}
      \caption{Estimation and monitoring profile of $T_L$}
      \label{fig:tlmonitor}
    \end{subfigure}
    \hfill
    \begin{subfigure}[b]{0.45\textwidth}
      \centering
      \includegraphics[width=\textwidth]{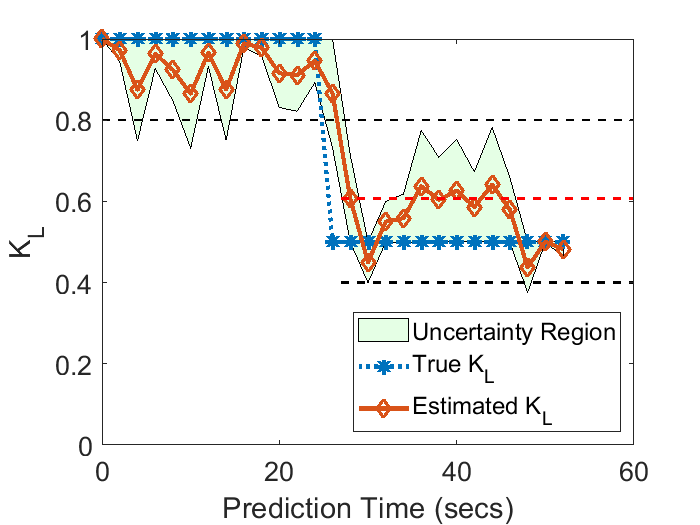}
      \caption{Estimation and monitoring profile of $K_L$}
      \label{fig:klmonitor}
    \end{subfigure}
\caption{Parameter monitoring and estimation}
\label{fig:monitoring}
\end{figure}

For the experimental setup shown above, we use our developed model to monitor these parameters. Specifically, our model collects data for the duration of 2 seconds (i.e., 200 data points with 0.01 time step) and then estimates the value of the parameters along with the respective uncertainty region. Fig. \ref{fig:monitoring} shows the real-time monitoring profiles of the parameters $T_L$ and $K_L$. We see from the figure that our framework can capture the change in these parameters and thus alert the controller. We note that the noise is injected at $Time=26$; however since we set the estimation interval at 2 seconds, there is a 2-second delay in the alert. Further, prior to noise injection, the estimates of the parameters were $T_L = 0.24$ and $K_L = 0.98$. After noise injection, the estimates were $T_L = 1.33$ and $K_L = 0.64$, closely resembling the true values.

In efforts to limit frequent and unnecessary changes in the parameters, we assign the bounds beyond which we consider an anomaly has occurred, and adjustment to the lower-level parameters in the controller is desired. Note that there are several ways to design such bounds. For example, since we have a probability distribution around our estimates, one can use statistical divergence methods to check when the estimated posterior distribution is statistically different than a baseline distribution of these estimates (a baseline distribution can be built using one's technical knowledge or user-specific). Another way could be various methods of statistical control charts. While all these could prove effective, we adopt here a margin of accepted error approach for a demonstration purpose, which is easy to implement and more practical. Specifically, boundaries are created based on a simple formulation of $Estimate \mp Accepted Change$. In our example, the accepted changes are set as $0.2$ for $T_L$ and $0.15$ for $K_L$, respectively. One can directly ask here why assign some bound and not update the $T_L$ and $K_L$ values directly. The principal need of such boundaries is that we do not wish to constantly change the values of $T_L$  and $K_L$ in our system. Continuously changing these parameters at every prediction time can result in unintended consequences on the performance by affecting vehicle jerk, acceleration, and comfort. Thus, ideally, we do not want to intervene with the controller when it is not really necessary to do so. We leave the pre-defined controller to work as is, until we deem it necessary or beneficial to change. This stresses on the importance of having a decision-making structure in addressing uncertainties in AVs; allowing us to take strategic actions. Another arise here on how to assign the boundaries (i.e., the \emph{AcceptedChange}). This can be left to the designer to assign or tune.

\begin{figure}[!htb]
\centering
\includegraphics[width=0.8\linewidth]{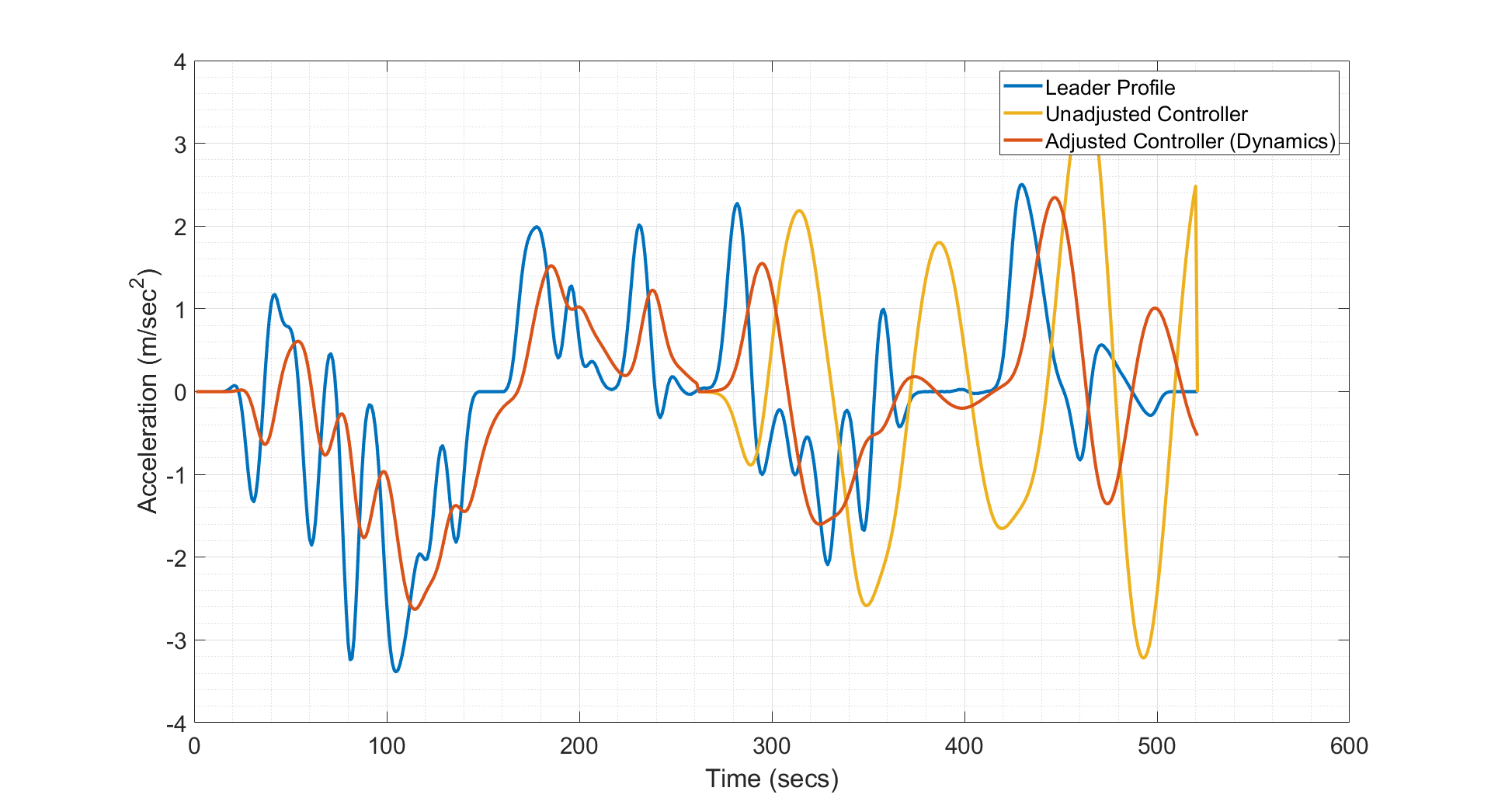}
\caption{Implications of adjusting lower-level settings based on real-time estimates on acceleration profile}
\label{fig:loweraccchange}
\end{figure}

\begin{figure}[!htb]
\centering
\includegraphics[width=0.8\linewidth]{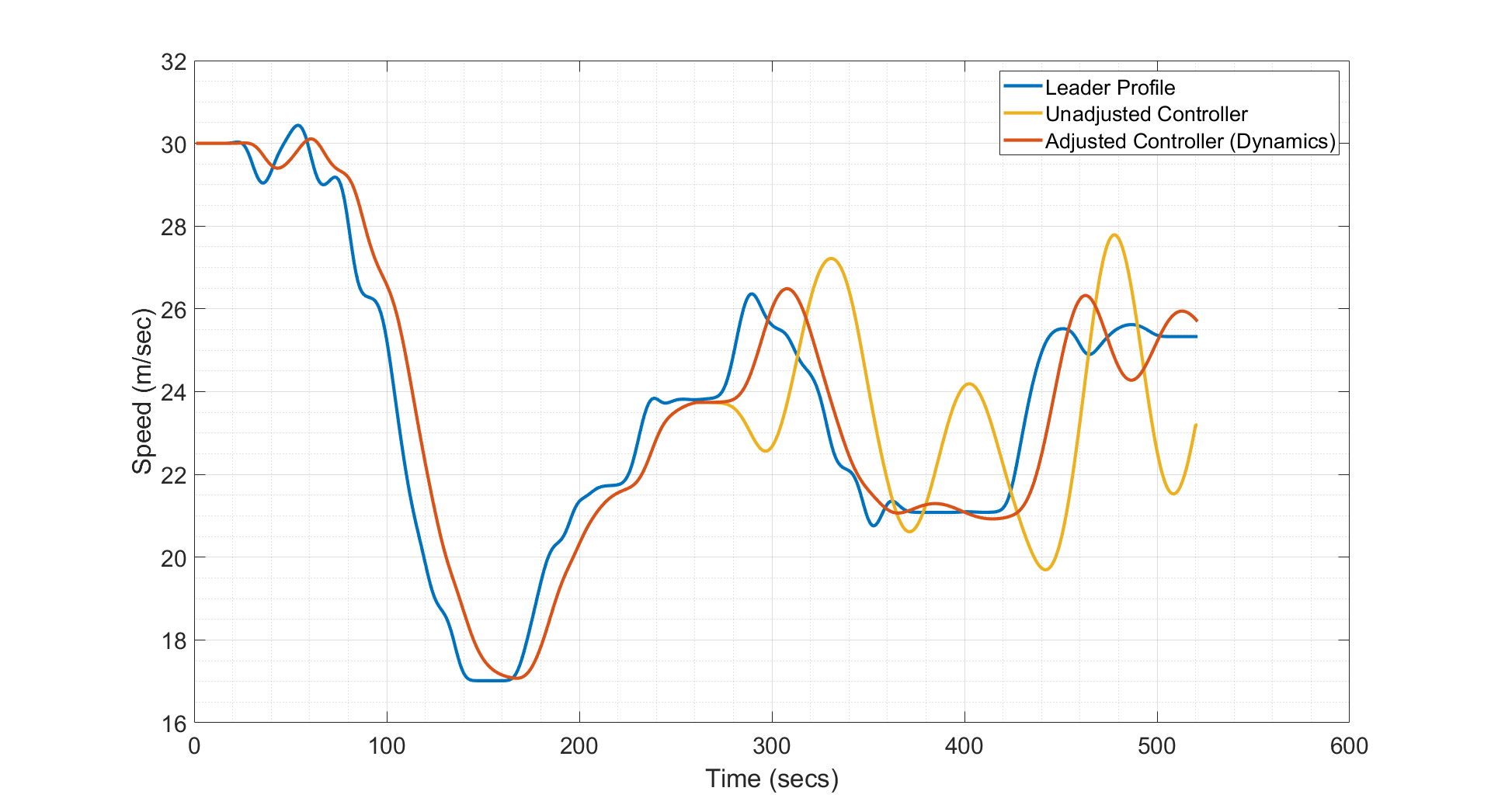}
\caption{Implications of adjusting lower-level settings based on real-time estimates on speed profile}
\label{fig:lowerspeedchange}
\end{figure}

Following Fig. \ref{fig:monitoring}, our goal now is to adjust the lower-level control setting for $T_L$ and $K_L$ to match the estimated ones from our model, $[T_L, K_L] = [1.33, 0.64]$. In our example, it is deemed critical to change the parameter set at $T=28$ (as they have crossed the accepted boundaries). Figs. \ref{fig:loweraccchange} \& \ref{fig:lowerspeedchange} show the acceleration and speed profiles, respectively, with adjustment (red line) and without (yellow). It is evident that the adjusted controller shows more stable behavior.

We note here that Strategy 1 also entails continuously monitoring the stability conditions in Eqs. 2 \& 3 (by plugging in any new estimates of $T_L$ \& $K_L$). In the case that these conditions are violated, this holds precedence over the boundary analysis, and more drastic action is taken, as introduced in the following sections. 

\subsubsection{Strategy 2: Time Gap Adjustment in Upper-Level Control}

\begin{figure}[!htb]
\centering
\includegraphics[width=0.8\linewidth]{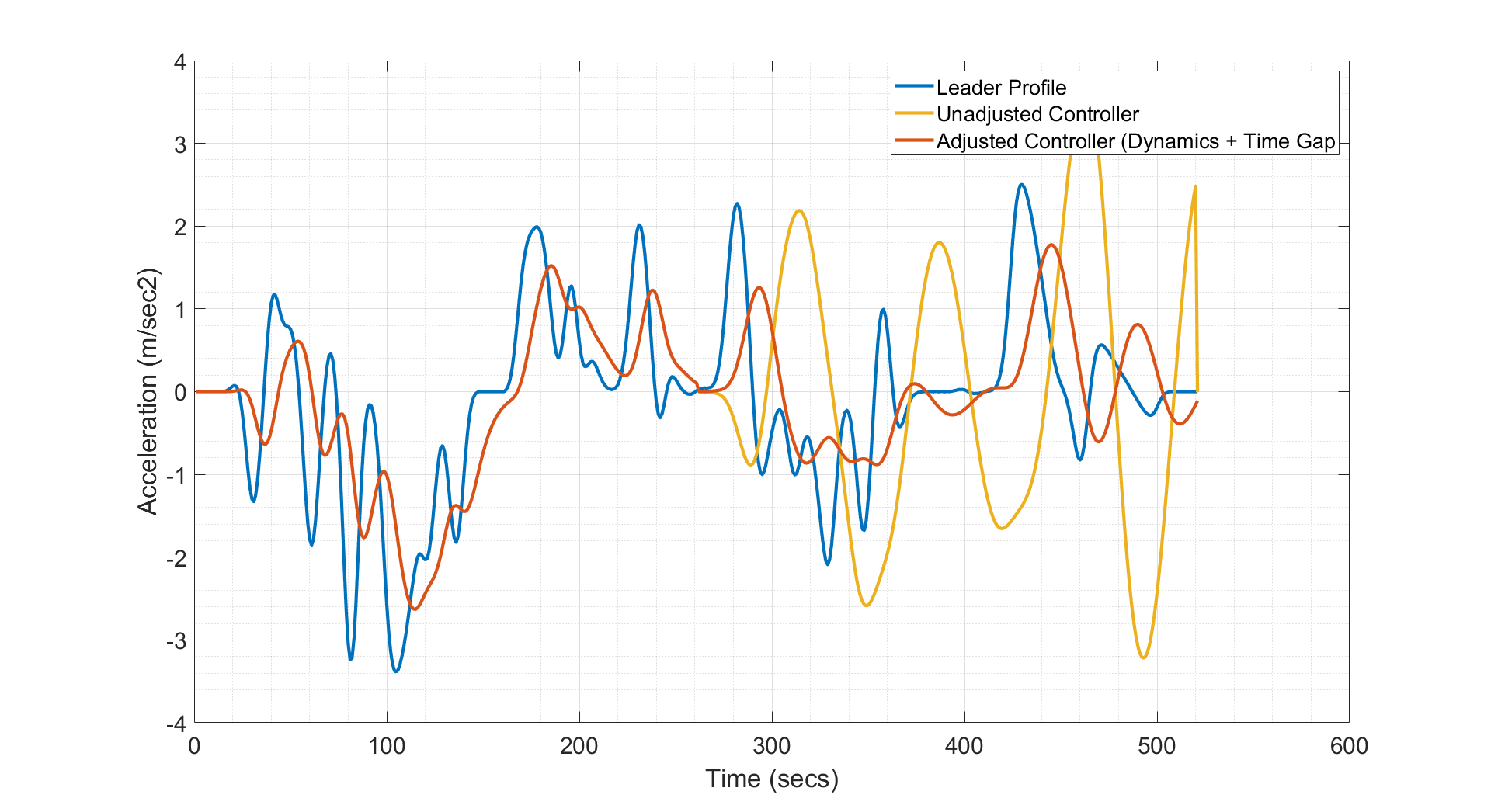}
\caption{Implications of adjusting lower-level settings and time gap based on real-time estimates - acceleration profile}
\label{fig:timegapchange}
\end{figure}

\begin{figure}[!htb]
\centering
\includegraphics[width=0.8\linewidth]{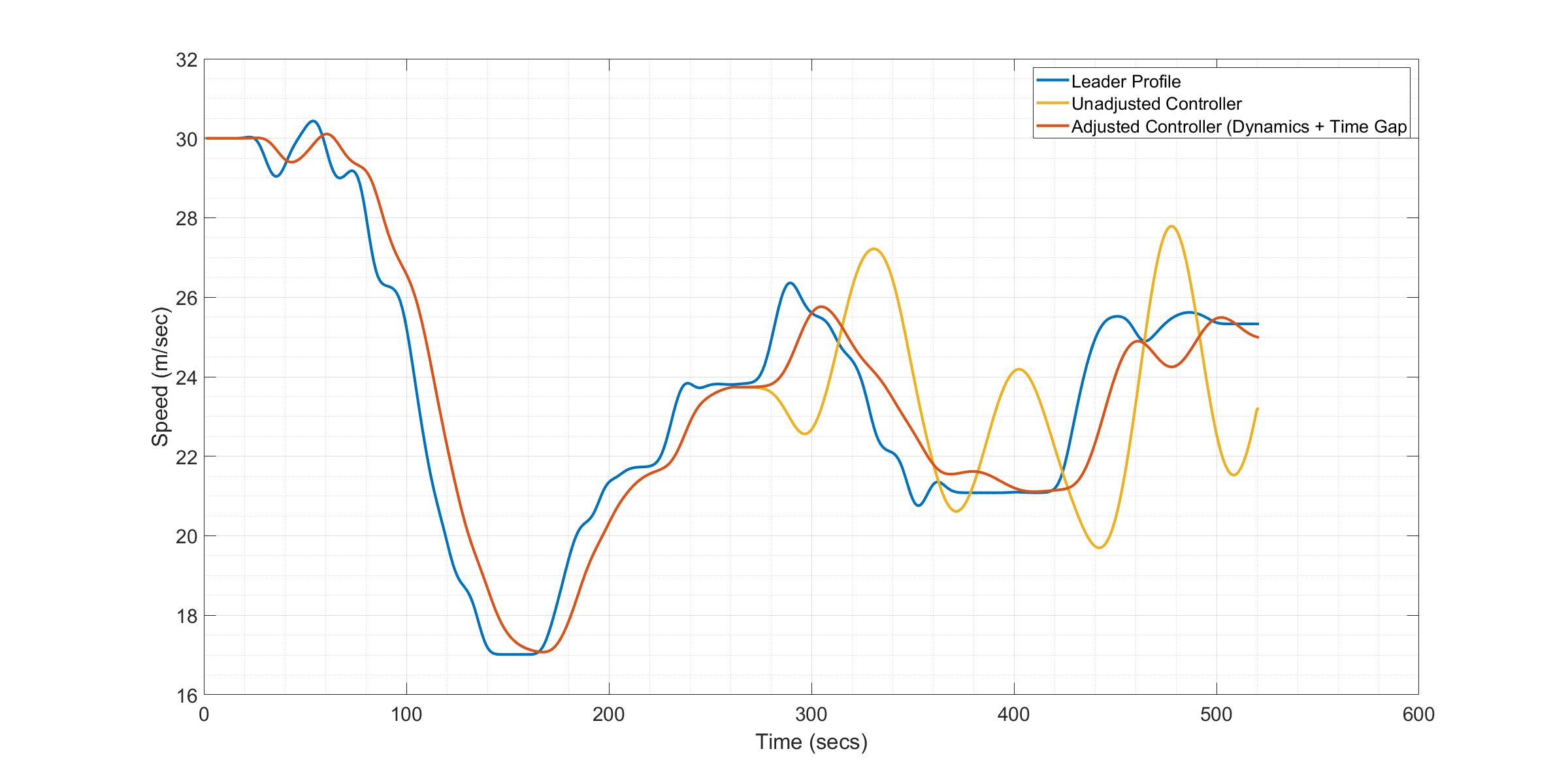}
\caption{Implications of adjusting lower-level settings and time gap based on real-time estimates - speed profile}
\label{fig:timegapchange_speed}
\end{figure}

It is possible that a vehicle encounters a situation that requires a more drastic action than updating the parameter values. One such situation is when we change the lower-level parameters and update them in our control structure, but still, we fall outside or on the border of the stability region defined by Eq. 2 \& 3. For example, in our experiment above, a more drastic action (other than $T_L$, $K_L$ update) is needed. As such, another strategy that complements strategy 1 is to adjust the time gap ($\tau^*$) in addition to changing $T_L$ and $K_L$. In our example, we increase $\tau^* = 2$, and adjust the lower-level parameters. The result of adopting this strategy is shown in Fig. \ref{fig:timegapchange} (acceleration profile) and Fig. \ref{fig:timegapchange_speed} (speed profile). The result suggests that changing the time gap is a very effective action in regaining stability and dampening the disturbance.

We note here that changing the time gap $\tau^*$ parameter can have a traffic level functionality. In essence, time gap is directly related to the traffic state by impacting roadway density, throughput, and traffic performance. Additionally, the time gap factor in a driver-specified metric in most ACC commercial vehicles available, thus changing it might have some implications on driver comfort, as it overrides what the driver desires. We further discuss these implications in Sec. \ref{S:4.5}

\newpage
\subsubsection{Strategy 3: Adjustment of Control Gains in Upper-Level Control}

In this strategy, we consider another adjustment in upper-level control. Specifically, we explore adjusting control gains $k = [k_s, k_v, k_a]$. In our previous work \citep{kontar2021multi}, we reveal the exact mechanisms by which the control gains affect the vehicle behavior and impact stability. In general, it was noted that an increase in control gains will prompt the vehicle to respond more efficiently to disturbances and thus help gain stability. However, the impact of each control gain is different, and thus this type of adjustment can be very complex. We summarize the general effect of these gains on the CF behavior in Table \ref{tab:effect}. For instance, an increase in $|k_s|$ or $|k_v|$ will prompt the controller to dampen disturbances and regain stability. However, the increase of $k_v$ is much more impactful than that of $k_s$, as it leads to a more responsive controller, thus, better disturbance dissipation. An interesting observation is that while increasing the absolute magnitude of $K_a$ makes the controller more resistant to acceleration change and thus less responsive. It can also make the controller less responsive to noise in acceleration change. This noise in acceleration can be generated due to uncertain vehicle dynamics. In another word, being less responsive to acceleration change can enhance the performance of your vehicle in scenarios where there exists high stochasticity (from exogenous and endogenous factors) in vehicular dynamics. Yet, this remains a complex behavior in need of further study.

\begin{table}[!htb]
\caption{Governing behavior of the control gains in linear controller, Reprinted from \cite{kontar2021multi}}
\centering
\scalebox{0.9}{
    \begin{tabular}{||l| l| l| l||}
    \hline
    \textbf{\bm{$k_i$}} & \textbf{Coefficient} & \textbf{Controller Command} & \textbf{Effect of $|k_i|$$\uparrow$}\\
    \hline
        $k_s$ & $\Delta d_i(t)$ & Maintain the target spacing & Pushes towards Neutral behavior \\
        $k_v$ & $\Delta v_i(t)$ & Match the leader's speed & Generates responsive behavior \\ 
        $k_a$ & $a_i(t)$ & Minimize acceleration & Resists acceleration change   \\
    \hline
\end{tabular}
}
\label{tab:effect}
\end{table}

\begin{figure}[!htb]
\centering
\includegraphics[width=0.8\linewidth]{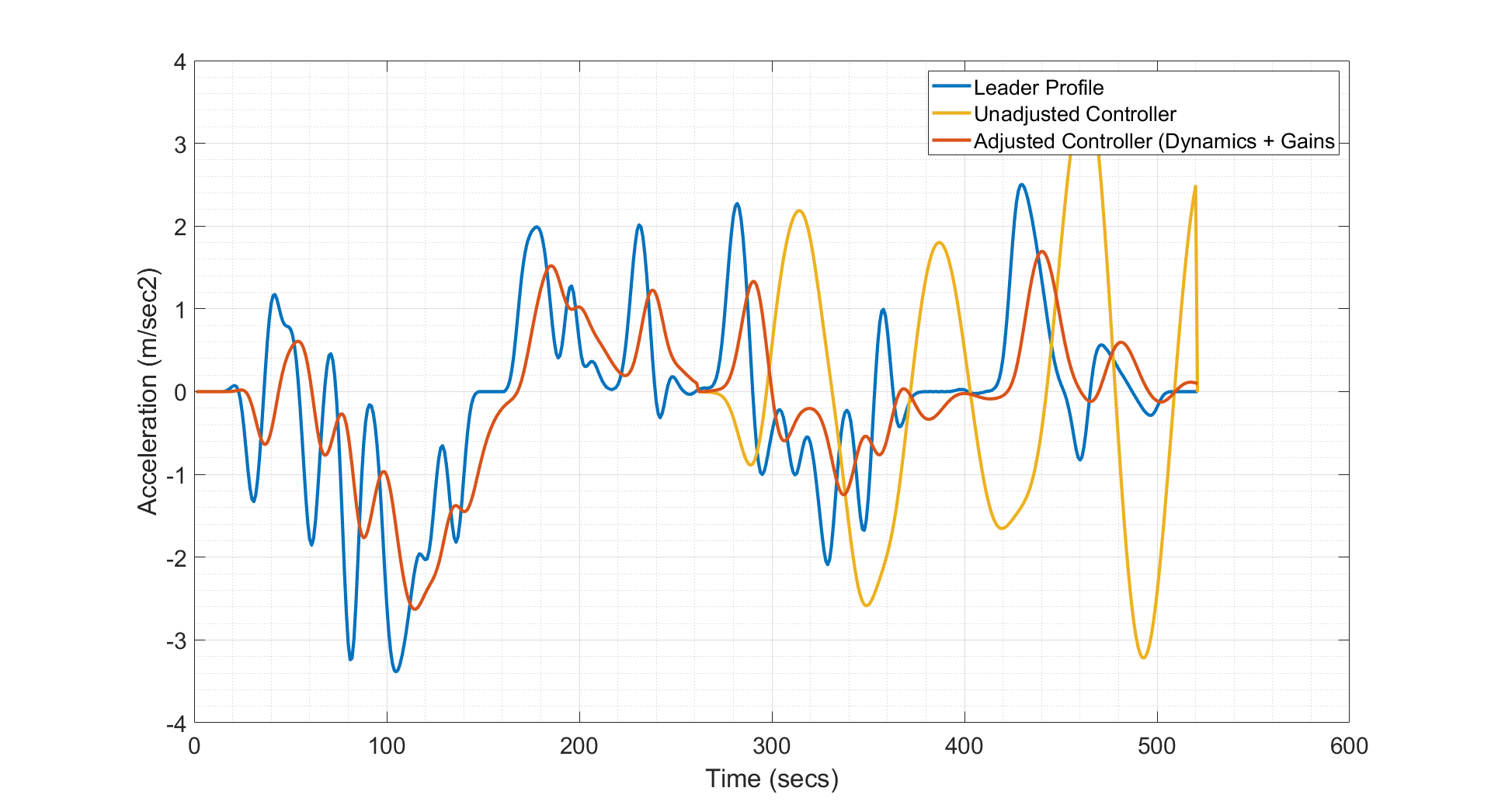}
\caption{Implications of adjusting lower-level settings and controller gains based on real-time estimates - acceleration profile}
\label{fig:changegains}
\end{figure}

\begin{figure}[!htb]
\centering
\includegraphics[width=0.8\linewidth]{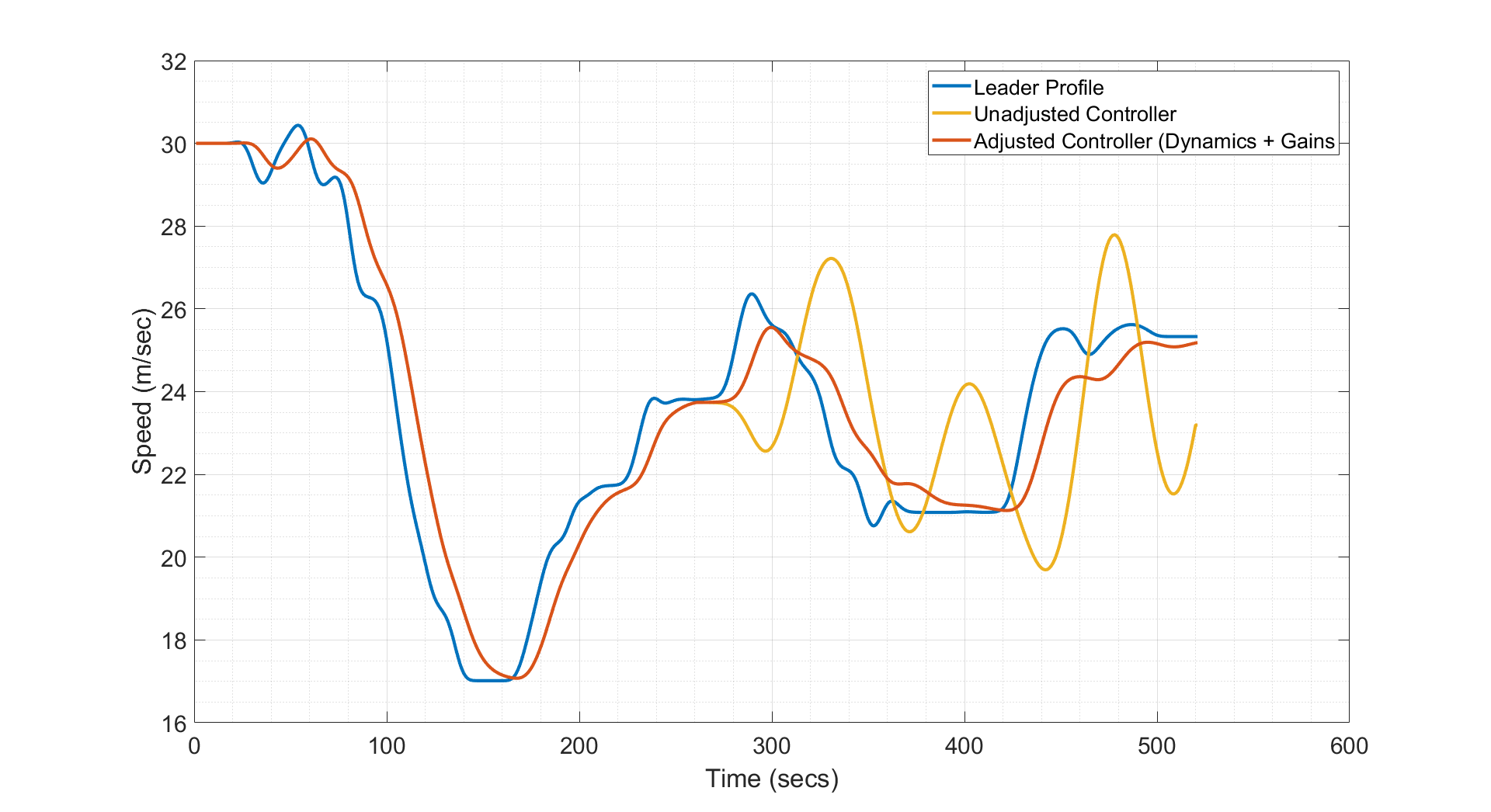}
\caption{Implications of adjusting lower-level settings and controller gains based on real-time estimates - speed profile}
\label{fig:changegains_speed}
\end{figure}

It is challenging to state precisely what values one should assign for the gains if they opt to adjust them. As mentioned before, the response is not linear and can get complicated. Delving into the details of this is not the focus of the paper, and we refer readers to our previous paper \citep{kontar2021multi} to gain a deeper understanding on this. However, the general principle here is to increase the gains in a way that stability conditions (Eq. 2 \& 3 for example) are satisfied (generally making your controller more responsive). For our experiment, we adjust the controller gains from $k=[1.5,1.5,-0.8]$ to $k=[3,3,-1.8]$, as well as adjusting the lower-level parameters (strategy 1). Results are shown in Fig. \ref{fig:changegains} (acceleration profile) and Fig. \ref{fig:changegains_speed}. It is evident from the profiles that stability is regained. 

\begin{algorithm}[!h]
\SetAlgoLined
\caption{Strategies Pseudo Code} 
\label{algo:algorithm2}

Input: Parameter estimates $\theta^{(t)} = \left[K^{(t)}_L, T^{(t)}_L\right]$\;

\If{$\theta^{(t)}$ outside acceptable boundary of $\left[K_L, T_L\right]$}
{Check Stability (Local and string)

\If{Stable}{Update lower-level controller: $\left[K_L, T_L\right] \leftarrow \left[K_L^{(t)},T_L^{(t)}\right]$}

\If{Unstable}{
Update lower-level controller $\left[K_L, T_L\right] \leftarrow \left[K_L^{(t)},T_L^{(t)}\right]$, and choose a strategy\;
\begin{enumerate}
    \item Update time gap: $\tau_{new}^* > \tau^* \leftarrow \tau^*$
    \item  Update Control gains: $\left[|k_s^{'}|>|k_s|, |k_v^{'}|>|k_v|, |k_a^{'}|>|k_a|\right]^{\top} \leftarrow \left[k_s, k_v, k_a\right]^{\top} $
\end{enumerate}
}}
\end{algorithm}

We summarize the strategic approach described above through Algorithm 2. In essence, the envisioned framework in this work is the integration of Algorithms 1 \& 2 together, which ultimately can help AVs achieve the desired performance level in real-life with the presence of endogenous and exogenous uncertainties. However, we acknowledge that largely the strategies can be implemented in different ways. As such, we provide a deeper analysis and discussion into the strategy implementation in Sec. \ref{S:4.5}

\newpage
\subsection{How to Implement a Strategy: Discussions and Remarks}\label{S:4.5}

One can clearly note that the methodology provided in this work is malleable and leaves space for personalized design and implementation. For instance, in implementing strategies 1, 2, \& 3 one can choose from a range of actions and also specify the level of controller modifications they want to make. This is rather a desired and natural outcome of this work, as strategic actions for AVs are situational. Building on this, we provide here some remarks on each strategy and things to consider. 

\noindent\textbf{1. On changing $T_L$ \& $K_L$ parameters:} As discussed in Sec. 4.02., when we continuously estimate $T_L$ and $K_L$ at each prediction time, it is not necessary that we update those parameters in the control system with the same frequency. The strategy here is to monitor these values and only change when the necessity arises. Interfering with the control system very frequently can lead to unintended performance consequences. We thus can rely on the monitoring methodology to alert us when an anomaly in $T_L$ and $K_L$ is triggered, and we can take the appropriate action afterward. Naturally, one can change the sensitivity of the monitoring methodology to anomaly detection (i.e., changing lower/upper bounds discussed in Fig. \ref{fig:monitoring}). This could be useful in cases where designers are demanding a very specific performance from an AV CF control and would not allow for any deviations. 

\noindent\textbf{2. On changing the time gap setting $\tau^*$:} Generally, when an anomaly is detected and we want to take action to regain stability of the AV, we would want to increase the time gap value. This prompts the CF controller to increase its separation distance from its follower and decreases the intensity of traffic disturbances when they reach the AV. The exact value of the increase in $\tau^*$ is hard to determine. Probably, the easiest and straightforward way to change $\tau^*$ is to stick to the available $\tau^*$ profiles in the vehicle. For example, in current commercial ACC vehicles, the driver can adjust the timegap parameter through a button on their dashboard. Usually, there are two-to-three settings: 1 second - 1.6 seconds - 2.5 seconds (note this can change from one vehicle model to another). Thus, when we want to increase $\tau^*$, we can alert the driver in situ to change their setting to the highest one. It is important to note here, that promoting the driver to change their desired timegap setting, can raise problems in driver compliance and comfort. Another important factor concerning the timegap setting is its relation to traffic level performance. 

Increasing timegap can decrease roadway density, throughput, and overall speed. However, it can help dissipate traffic disturbances. This aspect introduces a new dimension of our decision-making strategy, one that its not only related to enhancing traffic stability but also changing the traffic state and performance in real-time. 

\noindent\textbf{3. On changing gain setting $[k_s,k_v,k_a]$:} Changing the gain settings can be complex, as there exists a wide range of values that could have a similar impact. Generally, in adopting this strategy we aim at making the controller more responsive. However, the impact of the gain value is multifaceted. $k_s$ (gain for spacing) is neutral in behavior and does not yield a very responsive controller. $k_v$ (gain for speed) helps resolve disturbances fast by inducing a very responsive behavior. $k_a$ (gain for acceleration) resists acceleration and can oppose $k_v$. Thus, for this strategy, it might be most effective to just increase the values of $k_s$ and $k_v$. 

\noindent\textbf{4. Other Remarks (1):} Our methodology here is mostly data-driven. While we focus on the GLVD parameters, it is possible to use the same methodology in estimating real-time uncertainties for different parameters pertaining to AV performance. Additionally, it is possible to use this methodology on different type of controllers (i.e., Model Predictive Controller (MPC, reinforecement learning, etc.)

\noindent\textbf{5. Other Remarks (2):} Different strategies, other than timegap and controller gains, can be added to the general methodology of this work. Some notable ones could be monitoring communication delays, adjusting feedforward gains (if communication is present).

\color{black}
\section{Conclusions}\label{S:5}
This paper developed a methodological approach aimed at characterizing uncertainties in vehicular dynamics in an effort to monitor the AV car-following performance in real-time and support strategic actions to adjust control as needed. Towards this end, we adopted a Bayesian optimization scheme that reflects the stochastic nature of vehicular dynamics. The SGLD was utilized to solve the Bayesian optimization, allowing for uncertainty quantification in real-time. Building on this, we provided a new layer to AV CF control that enables continuous monitoring of controller performance and intervention if anomalies are detected. The developed methodology ultimately helps AVs achieve desired performance in presence of complex endogenous and exogenous uncertainties.

This study serves the need to design controllers that safeguards the vehicle's performance against real-time uncertainties. Nevertheless, several directions for future research is needed. This work focused only on vehicular dynamics uncertainties pertaining to acceleration behavior and could be expanded to study different types of uncertainties (e.g., first and second-order lags relative to location and speed of vehicle). We note here that our developed model is primarily data-driven and thus can easily be reproduced for different applications of uncertainty quantification in real-time. Additionally, in this work we provide decision-making strategies that can help maintain the desired performance of the controller specifically from stability point of view, however different strategies can be implemented to cover other performance metrics (e.g., comfort, vehicle efficiency, safe spacing, etc...). Finally, another important direction is to integrate such framework for controller of unknown structure (e.g., AI-based controller). These are part of ongoing research by the authors. 

\section*{Acknowledgment}
This research was sponsored by the United States National Science Foundation (NSF) through Award CMMI 1932932. 
\section*{Author Contribution Statement}
The authors confirm the contribution to the paper as follows: study conception and design: Kontar (lead) and Ahn; data collection: N/A; analysis and interpretation of results: Kontar (lead) and Ahn; draft manuscript preparation: Kontar (lead) and Ahn. All authors reviewed the results and approved the final version of the manuscript.

%%%%%%%%%%%%%%%%%%%%%%%%%%%%%%%%%%%%%%%%%%%%%%%%%%%%%%%%%%%%%%%%%%%%%%%%%%%%%%%%%%%%%
\bibliographystyle{elsarticle-harv}
\biboptions{authoryear}
\bibliography{references.bib}
%%%%%%%%%%%%%%%%%%%%%%%%%%%%%%%%%%%%%%%%%%%%%%%%%%%%%%%%%%%%%%%%%%%%%%%%%%%%%%%%%%%%%
\end{document}